\documentclass[12pt]{article}
\usepackage[dvipdfmx,hiresbb]{graphics}
\usepackage{float}
\usepackage{amsmath,amssymb}
\usepackage{mathrsfs}
\usepackage{braket}
\usepackage{framed,color}
\usepackage{graphicx}
\usepackage{bm}
\usepackage{cite}

\usepackage[utf8]{inputenc}
\usepackage{graphicx} 
\usepackage{color}
\usepackage{amsmath,amssymb}
\usepackage{amsthm}
\usepackage{bm,mathrsfs}
\usepackage{ulem}
\usepackage{float}
\usepackage{braket}
\usepackage{tcolorbox}
\usepackage[hang,small,bf]{caption}
\usepackage[subrefformat=parens]{subcaption}
\newcommand{\ten}[3]{{#1^{#2}}_{#3}}
\newcommand{\eq}[1]{$(\ref{eq#1})$}
\newcommand{\fig}[1]{$\ref{fig#1}$}
\newcommand{\tab}[1]{$\ref{tab#1}$}

\newcommand{\bras}[1]{\left( {#1} \right)}
\newcommand{\bram}[1]{\left\{ {#1} \right\}}
\newcommand{\brab}[1]{\left[ {#1} \right]}

\makeatletter

\renewcommand\section{\@startsection {section}{1}{\z@}%
                                   {-3.5ex \@plus -1ex \@minus -.2ex}%
                                   {2.3ex \@plus.2ex}%
                                   {\normalfont\large\bfseries}}

\renewcommand\subsection{\@startsection{subsection}{2}{\z@}%
                                     {-3.25ex\@plus -1ex \@minus -.2ex}%
                                     {1.5ex \@plus .2ex}%
                                     {\normalfont\normalsize\bfseries}}

\makeatother

\usepackage[driverfallback=dvipdfmx]{hyperref}
\hypersetup{
setpagesize=false,
 bookmarksnumbered=true,%
 bookmarksopen=true,%
 colorlinks=true,%
 linkcolor=blue,
 citecolor=blue,
}

\begin{document}

\baselineskip=18pt  
\numberwithin{equation}{section}  
\allowdisplaybreaks  



%
%


\thispagestyle{empty}

\vspace*{-2cm}
\begin{flushright}
\end{flushright}

\begin{flushright}
\end{flushright}

\begin{center}

\vspace{1.4cm}

{\bf \Large Decay of Kaluza-Klein Vacuum  }
\vspace*{0.2cm}

{\bf\Large  via Singular Instanton}

\vspace{1.3cm}

{\bf
Yutaka Ookouchi$^{1,2}$, Ryota Sato$^{2}$ and Sohei Tsukahara$^{2}$} \\
\vspace*{0.5cm}

${ }^{1}${\it Faculty of Arts and Science, Kyushu University, Fukuoka 819-0395, Japan  }\\
${ }^{2}${\it Department of Physics, Kyushu University, Fukuoka 810-8581, Japan  }\\

\vspace*{0.5cm}

\vspace*{0.5cm}

\end{center}

\vspace{1cm} \centerline{\bf Abstract} \vspace*{0.5cm}

In the decay process of metastable vacua in quantum field theories, the bounce solution, a classical solution in Euclideanized theories, is helpful in calculating the decay rate. Recently, the bounce solution with a conical singularity has attracted wide attention and revealed physical importance. In this paper, we discuss the bubble of nothing solution, which describes the decay process of a five-dimensional Kaluza-Klein vacuum, and study the consequence of including conical singularity. We found that the bounce solution with singularities has a higher decay rate than those without. This effect suggests that a singular solution can play a dominant role in vacuum decay of theories with compact internal space. We also discuss the enhanced decay rate from a thermodynamic perspective.

\newpage
\setcounter{page}{1} 

\newpage
\setcounter{page}{1} 



\section{Introduction}

The use of instantons for vacuum decay was first developed by Coleman and has become a powerful tool in quantum field theories \cite{Coleman:1977py,Callan:1977pt}. An analysis, including gravity, was conducted by Coleman and De Luccia \cite{Coleman:1980aw}, and various studies have been done since then; see Weinberg's textbook and references therein\cite{Weinberg}. 

An exciting development in this field is the analysis of vacuum decay using instantons with singularities. The study of singular instantons was initiated by Hawking and Turok \cite{Hawking:1998bn,Turok:1998he} and then explored further by many authors in the context of open universe\cite{Unruh:1998wc,Vilenkin:1998pp,Garriga:1998tm,Garriga:1998ri,Garriga:1998he,Hashimoto:2000zk}. In calculating decay rates using Euclideanized theories, it is not the metric that plays a physical role but the action. Hence, singularity contributions can have a physics role as long as they result in a finite value. Recently, regarding relatively well-behaved conical singularities, the authors of \cite{Gregory:2013hja} have developed the regularization method pioneered by Fursaev and Solodukhin\cite{Fursaev:1995ef} and applied it to calculate the bounce action, showing a specific prescription for the bounce solution with conical singularities. This method makes it easier to see that results are independent of the regularization method. 

In considering higher-dimensional theories, decay phenomena peculiar to compactified spacetime are particularly important. The simplest higher-dimensional spacetime is the Kaluza-Klein vacuum, which is a direct product of the four-dimensional Minkowski spacetime and a circle. In this vacuum, since the standard positive energy theorem\cite{Schon:1979rg,Schon:1981vd,Witten:1981mf} does not hold, then it will suffer from non-perturbative decay unless protected by any mechanisms such as introducing covariant fermions to the bulk. This is called ``{\it bubble of nothing}'' (BoN) first discussed by Witten \cite{Witten:1981gj}. In Witten's original paper, the period of imaginary time was chosen to a specific value and was well-tuned so that no singularity occurs where the size of the circle shrinks to zero, but there is room for reconsideration. As mentioned above, it is not the Euclidean metric that can have physical meaning, so if the contribution to the on-shell action is finite, the decay via singular instantons could be the dominant channel. Moreover, the bubble of nothing scenarios have been discussed mainly in the context of string theory\cite{Horowitz:2007pr,Blanco-Pillado:2016xvf,Acharya:2019mcu}, so it is worth investigating singular instantons in such scenarios in detail.\footnote{The possibility for relaxing the smoothness condition due to the existence of some regularization mechanism has also been mentioned in \cite{GarciaEtxebarria:2020xsr}.} 

In this paper, we relax the smoothness condition imposed on the bubble of nothing instanton in previous studies and consider the decay of the KK vacuum mediated by a singular instanton. While spacetime properties differ from those used in \cite{Gregory:2013hja}, the vicinity of the singularity is similarly characterized by a conical singularity, allowing us to calculate the bounce action based on the insightful method developed in \cite{Gregory:2013hja}. As we will explicitly demonstrate in this paper, the contribution of singularities significantly enhances the decay rate, thereby accelerating vacuum decay. This effect may provide an implication for the decay process in higher-dimensional theories with compact internal spaces. Our final goal is to gain insights into the vacuum structure of string theory. Towards this, using the concise model, we will examine the characteristic properties of singular instantons in higher-dimensional theories with vanishing internal spaces.

Our motivation for this research mainly comes from the ongoing Swampland Program \cite{Vafa:2005ui,Ooguri:2006in,Ooguri:2016pdq,Obied:2018sgi,Garg:2018reu,Ooguri:2018wrx,Lust:2019zwm} (see also \cite{Palti:2019pca,Brennan:2017rbf,Danielsson:2018ztv,Akrami:2018ylq,Grana:2021zvf,vanBeest:2021lhn} for reviews). In the Swampland Program, intriguing conjectures regarding the vacuum structure of string theory are being proposed. This program provides constraints on the low-energy effective theories of quantum gravity that can be derived from string theory. Among these, the de Sitter conjecture \cite{Obied:2018sgi,Garg:2018reu,Ooguri:2018wrx} proposes a new direction in creating our universe, and based on the arguments, many interesting models have been proposed, for example, by Danielson et al.\cite{Banerjee:2018qey,Banerjee:2019fzz,Banerjee:2020wix}. Even if all the conditions of the Swampland are met, a complex vacuum structure still can exist. If this is the case, it can be naturally believed that many transitions between metastable vacua occurred before the emergence of four-dimensional spacetime. In such vacuum transitions, our results presented in this paper may suggest that singular instantons can play a dominant role.

The organization of this paper is as follows: In Section 2, we quickly review the bounce solutions discussed by Witten \cite{Witten:1981gj} and introduce some notations and other preliminary necessities for the main calculations. Section 3 is the central part of this paper, where we evaluate the action for the bounce solution with a singularity. By dividing the vicinity of the singularity and the other parts, we calculate the bounce action basically along the lines of \cite{Gregory:2013hja}. As a result, we will find that the singular bounce solution gives a higher decay rate than that without. We also discuss the reason for the enhancement from a thermodynamic point of view in section 4. Section 5 is devoted to the discussion and outlook. In Appendix A, we summarize the notations used in this paper. In Appendix B, while clarifying the relationship with existing literature \cite{Gregory:2013hja}, we reproduce the calculations performed in this paper using the ADM decomposition.


\section{Quick review of bubble of nothing solution}
\label{Quick review of bubble of nothing solution}

In this section, we will briefly review the decay of the Kaluza-Klein vacuum introduced by Witten\cite{Witten:1981gj} and prepare for the calculations in the next section. The instanton solution utilized by Witten for the decay has $O(4)$ symmetry and asymptotically reach the Kaluza-Klein vacuum:
\begin{align}
    ds^2
    =
    \bras{1-\frac{\alpha}{r^2}}^{-1}dr^2+r^2d\Omega_3^2+\bras{1-\frac{\alpha}{r^2}}d\phi^2,
    \label{eq3.1.4}
\end{align}
where $d\Omega_3$ is the standard round metric. Since this solution seems to have a singularity at $r=\sqrt{\alpha}$, we define a new coordinate $\rho$ as  
\begin{align}
    \rho
    \equiv
    \sqrt{\alpha}\bras{1-\frac{\alpha}{r^2}}^{1/2},
    \label{eq3.1.5}
\end{align}
and examine the behavior near the would-be singularity in detail. Using this coordinate, we can approximate the instanton solution \eq{3.1.4} near the singularity as
\begin{align}
    ds^2
    \simeq
    d\rho^2 +\alpha^2d\Omega_3^2+ \rho^2d\bras{\frac{\phi}{\sqrt{\alpha}}}^2.
    \label{eq3.1.8}
\end{align}
In this coordinate system, the position of the singularity corresponds to $\rho=0$. In general, the period of $\frac{\phi}{\sqrt{\alpha}}$ is not $2\pi$, so there is a conical singularity at $\rho=0$ as shown in Figure \fig{3.2}, but if a smoothness condition $ \alpha  = R^2$ holds, the period of the variable $\phi$ is $2\pi R$ and there is no deficit angle. This is the nonsingular instanton solution introduced by Witten. This solution is called a bubble of nothing because the spacetime ends smoothly at $r =R$, and no spacetime exists inside it, as shown in the right panel in Figure \fig{3.2}. In the next section, we will examine singular solutions that do not satisfy the smoothness condition.

\begin{figure}[tb]
\centering
\includegraphics[width=50mm]{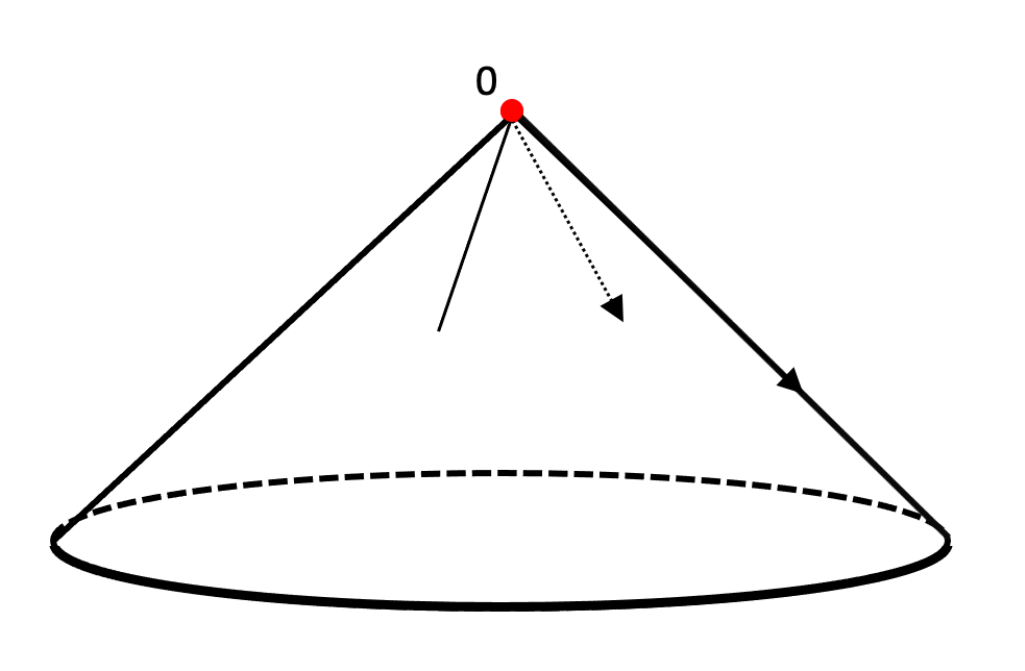} \hspace{2cm}
\includegraphics[width=70mm]{Sato33.pdf}
\caption{Schematic picture near the conical singularity of the instanton solution.}
\label{fig3.2}
\end{figure}

Now, let us calculate the bounce action substituting this solution into the Euclidean action. 5D gravitational action is 
\begin{align}
    I
    =
    -\frac{1}{16 \pi G_5} \int_{\mathscr V} \mathcal{R} \sqrt{g} d^5x
    -
    \frac{1}{8 \pi G_5} \oint_{\partial\mathscr V}
    \bras{\mathcal{K}-\mathcal{K}_{0}}\sqrt{\Tilde{g}}d^4\Tilde{x},
    \label{eq3.2.1}
\end{align}
where the first term is Einstein-Hilbert (EH) term, and the second is a boundary term called Gibbons-Hawking-York (GHY) term\cite{York:1972sj,Gibbons:1976ue}. $\mathcal{K}$ is the trace of extrinsic curvature and defined with outward unit normal vector $\tilde{n}_\alpha$ perpendicular to the surface as 
\begin{align}
    \mathcal{K}
    \equiv
    \ten{\Tilde{n}}{\alpha}{;\alpha}.
    \label{eq3.2.4}
\end{align}
Let us define the ``positive''ward unit normal vector $n_\alpha$ of the hypersurface for which $\Phi$ takes a constant value as
\begin{align}
    \ten{n}{}{\alpha}
    \equiv
    \frac{\partial_\alpha\Phi}{\sqrt{|\ten{g}{\mu\nu}{}\ten{\Phi}{}{,\mu}\ten{\Phi}{}{,\nu}|}}.
    \label{eq3.2.5}
\end{align}
$\Phi$ is a scalar value that characterizes the hypersurface.\footnote{For instance, $r$ constant hypersurface is characterized as 
\begin{equation*}
    \Phi
    =
    r-r_0
    =0,
\end{equation*}
where $r_{0}$ is an arbitrary constant.
}
Note that the outward unit normal $\tilde{n}_\alpha$ do not necessarily face the same direction as the unit normal vector $n_\alpha$ pointing in the positive direction of the coordinates. While the nondynamical term $\mathcal{K}_0$ does not affect the variation, this factor is needed to keep the Euclidean action finite.
$\ten{G}{}{5}$ is the five-dimensional gravitational constant and can be written with the compactification radius $R$ and the four-dimensional gravitational constant $G_4$ as $G_5=2\pi RG_4$. $\Tilde{x}$ is the coordinates of the boundary surface $\partial\mathscr V$, and we denote the induced metric on this surface as $\ten{\Tilde{g}}{}{ab}$.

Since the BoN instanton solution is the Euclidean 5D Schwarzshild solution and the scalar curvature vanishes,  the EH term does not contribute to calculating the on-shell action:
\begin{align}
    -\frac{1}{16 \pi G_5} \int_{\mathscr V} \mathcal{R} \sqrt{g} d^5x
    =
    0.
    \label{eq3.2.5.1}
\end{align}

\begin{figure}[tb]
\centering
\includegraphics[width=0.35\linewidth]{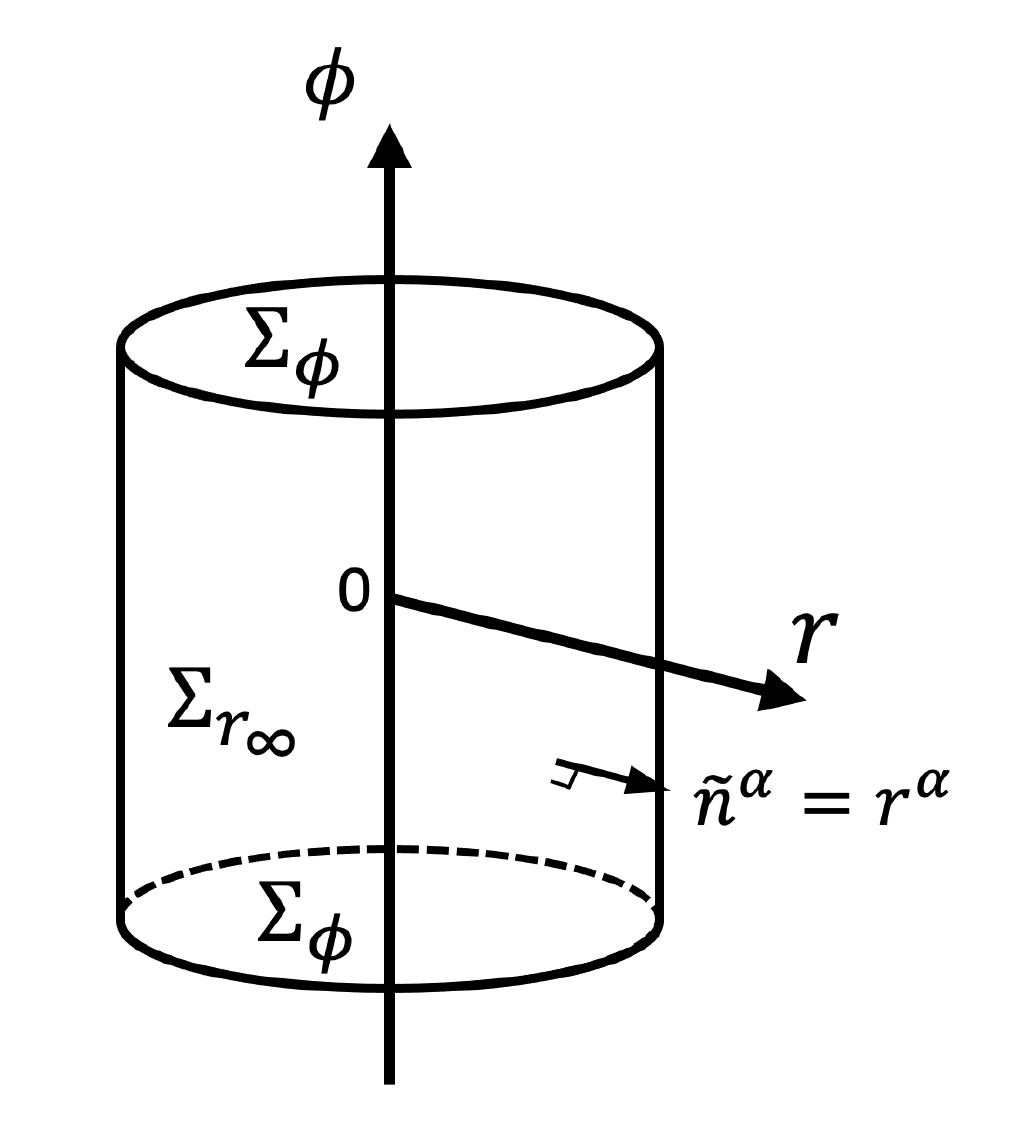}
\caption{Cylindrical boundary composed of $\Sigma_{\phi}$ and $\Sigma_{r}$.
}
\label{fig3.3}
\end{figure}

The next factor is the GHY term. Consider a cylinder composed of $\phi$-constant surfaces $\Sigma_\phi$ and an $r$-constant surface $\Sigma_{r}$ as shown in Figure \fig{3.3}. Note that since $\phi$ is the periodic coordinate, there is no actual boundary in the $\phi$ direction. Therefore, only the boundary integral on $\Sigma_{r}$ should be considered. That is
\begin{align}
        -\oint_{\partial\mathscr V}
        \bras{\mathcal{K}-\mathcal{K}_{0}}\sqrt{\Tilde{g}}d^4\Tilde{x}
        =
        \lim_{r_\infty\rightarrow{}\infty}\bras{-\int_{\Sigma_{r_\infty}}
        \bras{\mathscr{K}-\mathscr{K}_0}\sqrt{\gamma}d^4z},
    \label{eq3.2.6}
\end{align}
where $\int_{\Sigma_{r_\infty}}$ denotes the boundary integral on the asymptotic boundary $\Sigma_{r_\infty}$, and $\gamma$ is the determinant of the induced metric. See appendix \ref{notation} for our notation of geometric quantities. 
$\mathscr{K}$ is defined by a positiveward unit normal vector of an $r$-constant surface as 
\begin{align}
        \mathscr{K}
        \equiv
        r^\alpha_{;\alpha} .
    \label{eq3.2.6.1}
\end{align}
In this case, $\tilde{n}^\alpha$ and $r^\alpha$ are equal, so the sign before integration remains the same. Using $0\le \theta_{1,2}\le \pi$ and $0\le \theta_3\le 2\pi $, the three-dimensional round metric in the instanton solution can be expressed as
\begin{align}
    r^2d\Omega_3^2
    =
    r^2\bram{d\theta^{2}_{1}+\sin^2{\theta_1}
    (d\theta^{2}_{2}+\sin^2{\theta_2}d\theta^{2}_{3})} .
    \label{eq3.2.9}
\end{align}
Thus, the induced metric on $\Sigma_{r}$ is 
\begin{align}
    \gamma_{ij}dz^idz^j
    =
    r^2 d\theta^{2}_{1}+r^2\sin^2{\theta_1}d\theta^{2}_{2}+r^2\sin^2{\theta_1}\sin^2{\theta_2}d\theta^{2}_{3}+\bras{1-\frac{R^2}{r^{2}}}d\phi^2.
    \label{eq3.2.11}
\end{align}
This induced metric allows us to find the extrinsic curvature of the boundary terms, $\mathscr{K}$. Now, since the scalar quantity characterizing the hypersurface $\Sigma_{r_\infty}$ is expressed as $\Phi=r-r_\infty=0$, from \eq{3.2.5}, the positiveward unit normal vector is given by 
\begin{equation}
    \ten{r}{}{\alpha}
    =
    \frac{\partial_\alpha\bras{r-r_\infty}}{\sqrt{|\ten{g}{rr}{}\ten{\Phi}{}{,r}\ten{\Phi}{}{,r}|}}
    =
    \frac{\partial_\alpha r}{\sqrt{f(r)}}.
    \label{eq3.2.13}
\end{equation}
Consequently, we obtain the trace of the extrinsic curvature on $\Sigma_{r}$ as follow
\begin{align}
        \mathscr{K}
        = \ten{r}{\alpha}{;\alpha}
        = \frac{3}{r}\bras{1-\frac{R^2}{r^2}}^{\frac12}+\frac{R^2}{r^3}\bras{1-\frac{R^2}{r^2}}^{-\frac12}.
    \label{eq3.2.14}
\end{align}

Next, let us calculate the nondynamical term. Since $\mathscr{K}_0$ is the extrinsic curvature of hypersurfaces embedded into flat spacetime, we need to consider the embedding into flat spacetime represented by 
\begin{align}
    ds^2
    =
    d\tilde{r}^2+\tilde{r}^2d\tilde{\Omega}_3^2+d\tilde{\phi}^2.
    \label{eq3.2.15}
\end{align}
Specifically, the following relation for each coordinate is imposed.
\begin{align}
    \tilde{r} \equiv r,\quad
    \tilde{\Omega}_3 \equiv \Omega_3, \quad 
    d\tilde{\phi} \equiv \sqrt{1-\frac{R^2}{r^{2}}}d\phi,
    \label{eq3.2.15.1}
\end{align}
It can be seen that the flat spacetime \eq{3.2.15} matches the original instanton spacetime with $R=0$, so we obtain $\mathscr{K}_0$ simply by setting $R=0$ in \eq{3.2.14} as
\begin{align}
        \mathscr{K}_0
        =
        \frac{3}{\tilde{r}}
        = \frac{3}{r}.
    \label{eq3.2.16}
\end{align}
Then, using \eq{3.2.14} and \eq{3.2.16}, we can perform the boundary integral as 
\begin{align}
        -&\int_{\Sigma_{r_\infty}}
        \bras{\mathscr{K}-\mathscr{K}_0}\sqrt{\gamma}d^4z \notag
        \\&=
        -\int_{\Sigma_{r_\infty}}
        \bram{\frac{3}{r}\bras{1-\frac{R^2}{r^2}}^{\frac12}+\frac{R^2}{r^3}\bras{1-\frac{R^2}{r^2}}^{-\frac12}-\frac{3}{r}}r^3\sin^2{\theta_1}\sin{\theta_2}\sqrt{1-\frac{R^2}{r^{2}}}d^4z \notag
        \\&=
        -\int_{\Sigma_{r_\infty}}
        \bram{3r^2-2R^2-3r^2\bras{1-\frac{R^2}{r^2}}^{\frac12}}\sin^2{\theta_1}\sin{\theta_2}d^4z \notag \\
        & \simeq  -\int_{\Sigma_{r_\infty}}
        \bram{3r^2-2R^2-3r^2\bras{1-\frac{R^2}{2r^2}}}\sin^2{\theta_1}\sin{\theta_2}d^4z  \notag \\
        & = \int_{\Sigma_{r_\infty}}
        \frac{R^2}{2}\cdot\sin^2{\theta_1}\sin{\theta_2}d^4z \notag \\
        & = R^{2}\pi^{2}\cdot 2\pi R,
    \label{eq3.2.17}
\end{align}
where we used an approximation in the fourth line by taking the limit of $r_\infty\rightarrow\infty$ into account. Substituting this calculation into \eqref{eq3.2.6}, the on-shell value of the GHY term becomes
\begin{align}
        -\frac{1}{8 \pi G_5}\oint_{\partial\mathscr V}
        \bras{\mathcal{K}-\mathcal{K}_{0}}\sqrt{\Tilde{g}}d^4\Tilde{x}
        = \frac{\pi R^2}{8G_4}. \label{eq3.2.18.1}
\end{align}

Given the calculations \eq{3.2.5.1} and \eq{3.2.18.1}, the Euclidean action of the BoN instanton solution is
\begin{align}
        I
        \simeq
        \frac{\pi R^2}{8G_4} .
    \label{eq3.2.20}
\end{align}
The bounce action $B$ is defined as a difference between the Euclid action of the classical solution and the trivial solution. However, the latter is nothing but a flat vacuum solution, so its Euclidean action is obviously zero. Therefore, finally, we get\footnote{This result differs from Witten's original argument \cite{Witten:1981gj} by factor 2, but this discrepancy is confirmed by several studies \cite{Dowker:1995gb,Brown:2014rka}.
}
\begin{align}
        B
        \simeq
        I-I_0
        =
        \frac{\pi R^2}{8G_4} .
    \label{eq3.2.21}
\end{align}
Thus, we immediately see that the decay rate per unit volume and unit time is of the order of $\text{exp}(-\pi R^2/8G_{4})$.

Note that the above semiclassical analysis is valid only if the Kaluza-Klein radius is sufficiently large compared to the Planck length. Otherwise, the accuracy of approximation becomes poor, and the quantitative argument above becomes unreliable. However, even so, it can be strongly expected that the non-perturbative decay of the Kaluza-Klein vacuum would occur.

In this regard, we should mention that the compactification radius can not be determined at the classical level. A quantum correction will generally determine the radius, but the effective potential depends on which matter fields are present. In other words, the quantitative validity of this calculation is left to how the radius is fixed when quantum effects are included.

\section{Singular bubble of nothing solution}
\label{SingularBoN}

\subsection{Dividing of singular instanton spacetime}
\label{Dividing-BoN}

In the previous section, we considered the case where the instanton solution satisfies the smoothness condition, $\alpha =R^2$, following Witten's argument. In this section, we will relax this condition and deal with the metric with a conical singularity at $r=\sqrt{\alpha}$. Our basic argument follows \cite{Gregory:2013hja}. First, in order to calculate the bounce action of the BoN instanton solution with singularities, let us divide the instanton spacetime into two parts: the spacetime with singularities, $\mathscr{B}$, and the spacetime without singularities, $\mathscr{V}-\mathscr{B}$.

\begin{figure}[tb]
\centering
\includegraphics[width = 0.6\linewidth]{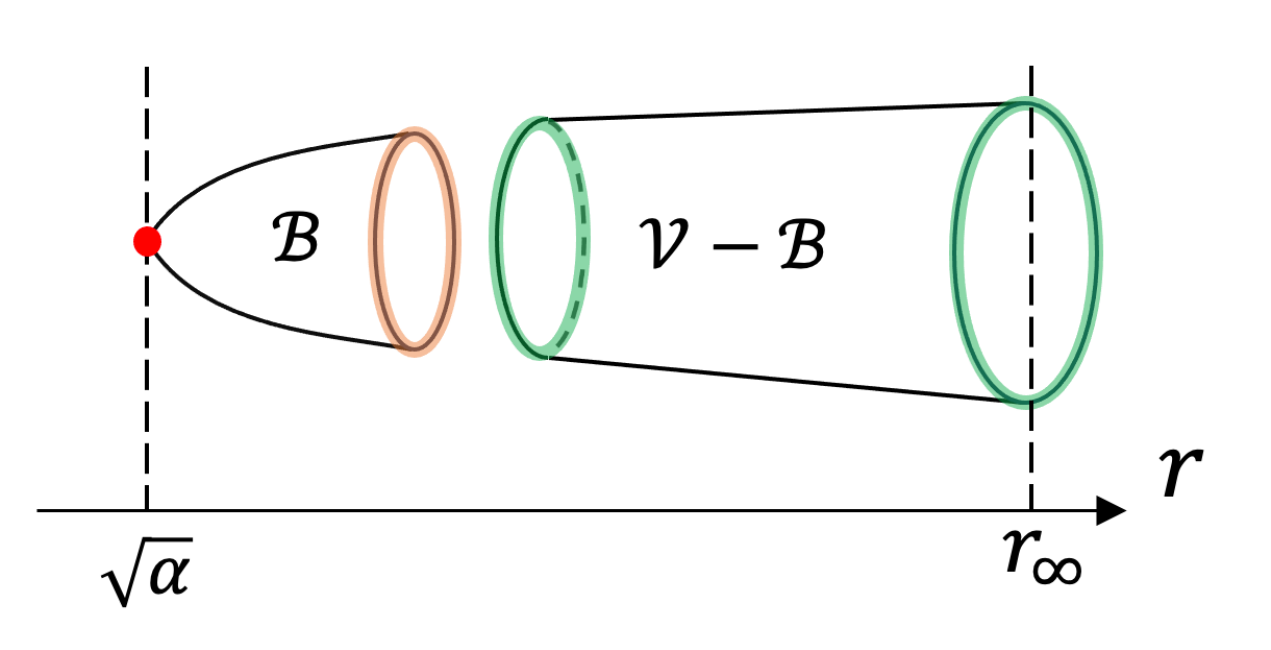}
\caption{The geometry of the BoN instanton solution split in two near the conical singularity represented by a red dot : The orange border is $\partial\mathscr{B}$ and the green border is $\partial \bras{\mathscr{V}-\mathscr{B}}$.}
\label{fig4.1}
\end{figure}
Accordingly, the Euclidean action of the 5D gravitational theory \eq{3.2.1} is also split into the Euclidean action near the singularity, $I_{\mathscr{B}}$, and the Euclidean action without the singularity, $I_{\mathscr{V}-\mathscr{B}}$. Namely,
\begin{gather}
    I
    =
    I_{\mathscr{V}-\mathscr{B}}+I_{\mathscr{B}},\label{eq4.1.3} \\
    I_{\mathscr{V}-\mathscr{B}}
    =
    -\frac{1}{16 \pi G_5} \int_{\mathscr{V}-\mathscr{B}} \mathcal{R}\sqrt{g} d^5x
    -\frac{1}{8 \pi G_5} \oint_{\partial \bras{\mathscr{V}-\mathscr{B}}} \bras{\mathcal{K}-\mathcal{K}_{0}}\sqrt{\Tilde{g}}d^4\Tilde{x},
    \label{eq4.1.4} \\
    I_{\mathscr{B}}
    =
    -\frac{1}{16 \pi G_5} \int_{\mathscr{B}} \mathcal{R}\sqrt{g} d^5x
    -\frac{1}{8 \pi G_5} \oint_{\partial \mathscr{B}} \bras{\mathcal{K}-\mathcal{K}_{0}}\sqrt{\Tilde{g}}d^4\Tilde{x}.
    \label{eq4.1.5}
\end{gather}
Note that the second terms of $I_{\mathscr{V}-\mathscr{B}}$ and $I_{\mathscr{B}}$ include contributions from not only the asymptotic boundary at infinity but also spatial boundaries that arise when spacetime is sliced off. In the subsequent sections, we will evaluate $I_{\mathscr{V}-\mathscr{B}}$ and $I_{\mathscr{B}}$ respectively in the limit where the location of the splitting is close to the singularity and finally sum them to obtain the total Euclidean action.\footnote{When dividing a spacetime that does not contain conical singularities, the integrals on the spatial boundaries contained in $\mathscr{V}-\mathscr{B}$ and $\mathscr{B}$ have the same value with different signs. On the other hand, when considering a spacetime that includes singularities, it is nontrivial that these terms coincide since there may be a special contribution from singularities. However, as can be seen from the calculations shown later, they coincide even in the presence of singularities.}

\subsection{Euclidean action of non-singular part}
\label{EuclidAction_V/B}
We will calculate the on-shell action of the manifold that does not have singularities.
\begin{figure}[tb]
\centering
\includegraphics[width=0.5\linewidth]{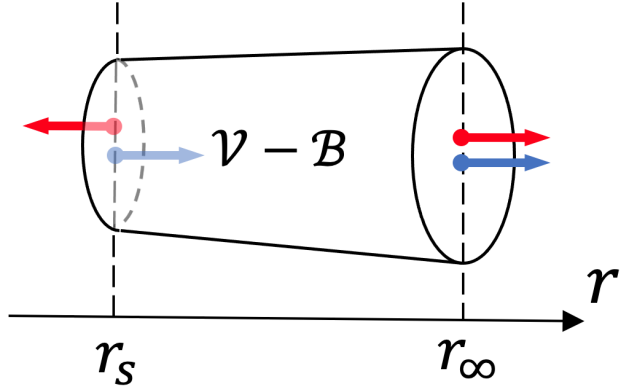}
\caption{Outward unit normal vectors $\Tilde{n}^\alpha$ represented by red allows and positiveward unit normal vectors $r^\alpha$ represented by blue allows.}
\label{fig4.2}
\end{figure}
Since the BoN instanton solution is Ricci flat except for the singularity, there is no contribution from the EH term. In contrast, for the GHY term, there are contributions from the spatial boundary $\Sigma_{r_s}$ and the asymptotic boundary $\Sigma_{r_\infty}$ as shown in Figure \fig{4.2}. In other words, to evaluate the on-shell value, we should concentrate on two boundary integrals
\begin{align}
	\begin{aligned}
        &-\oint_{\partial(\mathscr V-\mathscr B)}
        \bras{\mathcal{K}-\mathcal{K}_{0}}\sqrt{\Tilde{g}}d^4\Tilde{x} 
        \\&\hspace{1.5cm}=
        \lim_{r_s\rightarrow{}\sqrt{\alpha}}
        \int_{\Sigma_{r_s}}
        \bras{\mathscr{K}-\mathscr{K}_0}\sqrt{\gamma}d^4z
        -\lim_{r_\infty\rightarrow{}\infty}
        \int_{\Sigma_{r_\infty}}
        \bras{\mathscr{K}-\mathscr{K}_0}\sqrt{\gamma}d^4z,
        \end{aligned}
    \label{eq4.2.2}
\end{align}
where $\mathscr{K}$ is the trace of the extrinsic curvature defined by the positiveward unit normal vector $r^\alpha$ of the $r$-constant surface. Note that the relative sign of $\Tilde{n}^\alpha$ and $r^\alpha$ on $\Sigma_{r_s}$ and $\Sigma_{r_\infty}$ are
\begin{align}
	&\Tilde{n}^\alpha=-r^\alpha \quad \text{on $\Sigma_{r_s}$}, \\
	&\Tilde{n}^\alpha=r^\alpha \quad \text{on $\Sigma_{r_\infty}$}.
\end{align}
Thus, we need to pay attention to the fact that the sign of each boundary integral is different. The induced metric on $\Sigma_{r}$ is given by
\begin{align}
    \gamma_{ij}dz^idz^j
    =
    r^2 d\theta^{2}_{1}+r^2\sin^2{\theta_1}d\theta^{2}_{2}+r^2\sin^2{\theta_1}\sin^2{\theta_2}d\theta^{2}_{3}+\bras{1-\frac{\alpha}{r^{2}}}d\phi^2.
    \label{eq4.2.4}
\end{align}
Using this metric, we can derive the traces of extrinsic curvatures in boundary integrals. Since scalar quantities characterizing $\Sigma_{r_s}$ and $\Sigma_{r_\infty}$ are represented by $\Phi=r-r_s=0$ and $\Phi=r-r_\infty=0$, respectively, the positiveward unit normal vectors obtained from the definition \eq{3.2.5} are
\begin{align}
    \ten{r}{}{\alpha}^{(s)}
    =
    \frac{\partial_\alpha\bras{r-r_s}}{\sqrt{|\ten{g}{rr}{}\ten{\Phi}{}{,r}\ten{\Phi}{}{,r}|}}
    =
    \frac{\partial_\alpha r}{\sqrt{f(r)}}\ , \qquad 
    \ten{r}{}{\alpha}^{(\infty)}
    =
    \frac{\partial_\alpha\bras{r-r_\infty}}{\sqrt{|\ten{g}{rr}{}\ten{\Phi}{}{,r}\ten{\Phi}{}{,r}|}}
    =
    \frac{\partial_\alpha r}{\sqrt{f(r)}}.
    \label{eq4.2.7}
\end{align}
Clearly, the expression for $r_{\alpha}$ is the same for both surfaces. Therefore, the extrinsic curvature for both hypersurfaces is given by
\begin{align}
        \mathscr{K}
  =
        \ten{r}{\alpha}{;\alpha}
        =
        \ten{g}{\alpha\mu}{}\ten{r}{}{\mu;\alpha}
        =
        \frac{3}{r}\bras{1-\frac{\alpha}{r^2}}^{\frac12}+\frac{\alpha}{r^3}\bras{1-\frac{\alpha}{r^2}}^{-\frac12}.
    \label{eq4.2.8}
\end{align}

Next, we will consider the nondynamical term. $\mathscr{K}_0$ is the extrinsic curvature of the $r$-constant surface, which is embedded into the flat spacetime, and the embedding method is the same as described in section \ref{Quick review of bubble of nothing solution}. In the end, we need only to fix $\alpha$ to zero in \eq{4.2.8} to obtain
\begin{align}
        \mathscr{K}_0
        =
        \frac{3}{r}.
    \label{eq4.2.9}
\end{align}

Substituting \eq{4.2.8} and \eq{4.2.9} into \eqref{eq4.2.2}, we obtain
\begin{align}
        \int_{\Sigma_{r_s,r_\infty}}
        \bras{\mathscr{K}-\mathscr{K}_0}\sqrt{\gamma}d^4z
        =
        \int_{\Sigma_{r_s,r_\infty}}
        \bram{3r^2-2\alpha-3r^2\bras{1-\frac{\alpha}{r^2}}^{\frac12}}\sin^2{\theta_1}\sin{\theta_2}d^4z. \label{eq4.2.10}
\end{align}
This expression is exactly the same as the one replacing $R^2$ with $\alpha$ in the third line in \eq{3.2.17}. Finally, we will take limits of $r_s\rightarrow{}\sqrt{\alpha}$ and $r_\infty\rightarrow{}\infty$ respectively. For the integral on $\Sigma_{r_s}$, we can take the limit of $r_s\rightarrow{}\sqrt{\alpha}$ directly at \eq{4.2.10}, so the  calculation of the boundary integral becomes
\begin{equation}
        \lim_{r_s\rightarrow{}\sqrt{\alpha}}\int_{\Sigma_{r_s}}
        \bras{\mathscr{K}-\mathscr{K}_0}\sqrt{\gamma}d^4z
        =
        \int_{\Sigma_{\sqrt{\alpha}}}
        \alpha\sin^2{\theta_1}\sin{\theta_2}d^4z
        = 2\pi^{2} \alpha \cdot 2\pi R.
    \label{eq4.2.12}
\end{equation}
In contrast, the integral on $\Sigma_{r_\infty}$ results in
\begin{align}
    -\lim_{r_\infty\rightarrow{}\infty}\int_{\Sigma_{r_\infty}}
        \bras{\mathscr{K}-\mathscr{K}_0}\sqrt{\gamma}d^4z
        \simeq
        \lim_{r_\infty\rightarrow{}\infty}\int_{\Sigma_{r_\infty}}
        \frac{\alpha}{2}\cdot\sin^2{\theta_1}\sin{\theta_2}d^4z 
        = \frac{2\pi^{2}\alpha}{2} \cdot 2\pi R,\label{eq4.2.13}
\end{align}
where we took an approximation at large $r$ as in \eq{3.2.17}. Therefore, substituting \eq{4.2.12} and \eq{4.2.13} into \eq{4.2.2}, the contribution from GHY term is obtained as 
\begin{align}
        -\frac{1}{8 \pi G_5}&\oint_{\partial(\mathscr V-\mathscr B)}
        \bras{\mathcal{K}-\mathcal{K}_{0}}\sqrt{\Tilde{g}}d^4\Tilde{x}
        \simeq \frac{1}{8 \pi G_5} \cdot 2\pi^2\alpha \cdot 2\pi R +\frac{1}{8 \pi G_5} \cdot \frac{2\pi^{2}\alpha}{2} \cdot 2\pi R      
        = \frac{3\pi \alpha}{8G_4}. \label{eq4.2.13.1}
\end{align}
The first term in the middle expression comes from the spatial boundary resulting from dividing spacetime. Since the boundary integral takes a finite value,  this may seem to contradict the argument by the authors of \cite{Gregory:2013hja}. However, this difference is due to the presence of the asymptotic boundary in the BoN instanton solution, so our result is also consistent with their statement. See appendix \ref{EuclidAction-3+1} for details.

From the above discussion, the on-shell action of the spacetime with singularity cut off is as follows 
\begin{align}
        I_{\mathscr V-\mathscr B}
        \simeq
        \frac{3\pi \alpha}{8G_4}.
    \label{eq4.2.15}
\end{align}

\subsection{Euclidean action of singular part}
\label{EuclidAction_B}

In this section, we derive the Euclidean action of the spacetime with singularity, denoted as $I_{\mathscr{B}}$,  by conical deficit regularization\cite{Fursaev:1995ef,Gregory:2013hja}. Before regularization, let us introduce new coordinates and study their properties near the singularity:
\begin{equation}
    \rho
    \equiv
    r\sqrt{f}, \quad
  f = 1-\frac{\alpha}{r^2}.
    \label{eq4.3.1}
\end{equation}
In these coordinates, the instanton solution is expressed as follows
\begin{equation}
    ds^2
    =
    f(r(\rho))d\phi^2+d\rho^2+r(\rho)^2d\Omega^2_3.
    \label{eq4.3.2}
\end{equation}
When written in this coordinate system, the position of the conical singularity originally located at $r=\sqrt{\alpha}$ is expressed as $\rho=0$. Furthermore, replacing $\phi$ with a $2\pi$-periodic variable $\chi$, we can rewrite the instanton solution as follows
\begin{align}
    ds^2
    =
   F(\rho)^2d\chi^2+d\rho^2+r(\rho)^2d\Omega^2_3,\quad 
    F(\rho)^2 \equiv f(r(\rho))R^2,
    \label{eq4.3.3}
\end{align}
where we used
\begin{align}
    \frac{d\phi}{2\pi R}
    =
    \frac{d\chi}{2\pi}.
    \label{eq4.3.4}
\end{align}
To capture the behavior of the coordinates at a singularity, we need to perform a Taylor expansion of the coordinates in the vicinity of the singularity. From the definition of \eqref{eq4.3.3}, $F(0)=0$ holds at the location of the singularity, so the Taylor expansion of $F(\rho)$ near the location is
\begin{align}
    F(\rho)\simeq\rho F'(0),
    \label{eq4.3.6}
\end{align}
where the prime symbol (${}^{\prime}$) denotes the derivative with respect to $\rho$. Substituting \eq{4.3.6} into \eq{4.3.3}, we can confirm that the instanton solution near the singularity is approximately as follows
\begin{align}
    ds^2
    \simeq
    d\rho^2+\rho^2d(F'(0)\chi)^2+r(0)^2d\Omega^2_3.
    \label{eq4.3.7}
\end{align}
This equation shows that if $F^{\prime}(0)\chi$ has $2\pi$ periodicity, i.e. $F^{\prime}(0)=1$, there is no singularity at $\rho=0$. However, since $F^{\prime}(0)\neq1$ in general, there would be a deficit angle $2\pi\delta$ defined as
\begin{align}
    \delta
    =
    1-\frac{F(\rho)}{\rho}\bigg|_{\rho\xrightarrow{}0}
    =
    1-F'(0),
    \label{eq4.3.8}
\end{align}
where we used \eq{4.3.6} in the last deformation. Here, since the expression for $F(\rho)$ is already clear, its derivative can be calculated immediately. In fact, using \eq{4.3.1} and \eq{4.3.3}, we obtain
\begin{align}
    \delta
    =
    1-\frac{R}{\sqrt{\alpha}},
    \label{eq4.3.8.2}
\end{align}
as the specific expression for the deficit angle.

\begin{figure}[tb]
\centering
\includegraphics[width=0.5\linewidth]{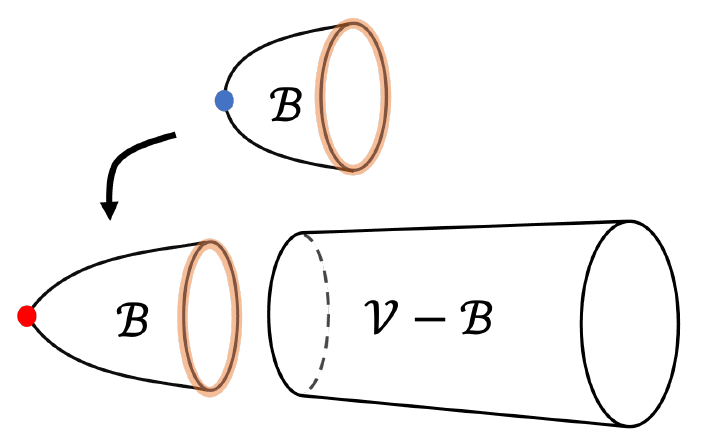}
\caption{Schematic picture of regularization procedure for conical singularity.}
\label{fig4.3}
\end{figure}

Based on the above analysis, let us perform regularization. Specifically, we will regularize by splitting the spacetime near the singularity and replacing it with a smooth manifold, as shown in Figure \fig{4.3}. In this paper, we adopt such a regularization without any proof, but such smoothing is expected to happen naturally in the framework of string theory. For example, in \cite{Adams:2001sv}, the authors argued that the conical singularity is eliminated by closed string tachyon condensation. If we consider the singularity as a sign that a new degree of freedom is missing, taking it into account appropriately, it would be regularized. However, it is important to note that the calculations here do not depend on the details of the regularization, and our conclusions do not change no matter how the regularization is performed.

We introduce a regularized metric as
\begin{align}
    ds^2
    =
    \tilde{F}(\rho)^2d\chi^2+d\rho^2+r(\rho)^2d\Omega^2_3,
    \label{eq4.3.9}
\end{align}
where the tilde-added function reflects the profile of regularized spacetime. In other words, $\tilde{F}^{\prime}(0)=1$ holds because this spacetime has no conical singularity. We denote the position of splitting as $\rho=\epsilon$, where $\epsilon$ is a tiny quantity. We will take a limit of $\epsilon\xrightarrow{}0$ in the last of our calculation. Moreover, this metric must match the original spacetime at $\rho = \epsilon$, so the deficit angle necessarily satisfies
$  \delta
    =    1-\frac{\tilde{F}(\epsilon)}{\epsilon}$.
Since $\epsilon$ is a very small quantity, this equation can be immediately rewritten as
\begin{align}
    \delta
    \simeq
    1-\tilde{F}'(\epsilon).
    \label{eq4.3.12}
\end{align}

Let us compute the Euclidean action under the regularization. The Ricci scalar for \eq{4.3.9} is calculated as 
\begin{align}
    \mathcal{R}
    =
    -\frac{2\tilde{F}''}{\tilde{F}}
    -\frac{6r''(\rho)}{r(\rho)}
    -\frac{6\tilde{F}'r'(\rho)}{\tilde{F}r(\rho)}
    -\frac{6(1-r'(\rho)^2)}{r(\rho)^2}.
    \label{eq4.3.13}
\end{align}
Among these terms, divergent terms in the limit of $\epsilon\xrightarrow{}0$ will be dominant for the calculation. By definition, order estimations of $\tilde{F}(\rho)$ and its second derivative yield
\begin{gather}
    \tilde{F}
    \simeq
    \epsilon\tilde{F}^{\prime}(0)
    =
    \mathcal{O}(\epsilon) ,
    \label{eq4.3.15} \\
    \tilde{F}^{\prime \prime}
    =
    \mathcal{O}\bras{\frac{\tilde{F}'(\epsilon)-\tilde{F}'(0)}{\epsilon-0}} 
    =
    \mathcal{O}\bras{\frac{(1-\delta)-1}{\epsilon}} 
    =
    \mathcal{O}\bras{-\frac{\delta}{\epsilon}}, 
    \label{eq4.3.14}
    \end{gather}
then we obtain
\begin{gather}
    -\frac{2\tilde{F}^{\prime \prime}}{\tilde{F}}
    =
    \frac{\mathcal{O}\bras{\frac{\delta}{\epsilon}}}{\mathcal{O}(\epsilon) }
    =
    \mathcal{O}\bras{\frac{\delta}{\epsilon^2}},
    \label{eq4.3.16} \\
    -\frac{\tilde{F}^{\prime}r^{\prime}}{\tilde{F}r}
    =
    \frac{\mathcal{O}(\epsilon)}{\mathcal{O}(\epsilon) }
    =
    \mathcal{O}\bras{\epsilon^0}.
    \label{eq4.3.17}
\end{gather}
From the above estimations, it can be seen that in the limit of $\epsilon\xrightarrow{}0$, only the first term of the Ricci scalar diverges and comes into effect in the action integral. Based on this, the EH term in the limit can be calculated as follows
\begin{align}
    -\frac{1}{16\pi G_5}\int_{\mathscr{B}}\mathcal{R}\sqrt{g} d^5x
    &=
    \lim_{\epsilon\rightarrow0}\bras{
    -\frac{1}{16\pi G_5}
    \int_0^\epsilon d\rho 
    \int_{\Sigma_\rho} d^4z
    \tilde{F}(\rho)r(\rho)^3\sin^2{\theta_1}\sin{\theta_2} 
    \mathcal{R}} \notag \\
    &\simeq
    \lim_{\epsilon\rightarrow0}\bram{
    -\frac{\pi^2}{4G_5}
    \int_0^\epsilon d\rho 
    \tilde{F}(\rho)r(\rho)^3
    \bras{-\frac{2\tilde{F}''}{\tilde{F}}}} \notag
    \\
    &\simeq
    \lim_{\epsilon\rightarrow0}\bras{
    \frac{\pi^2r(\epsilon)^3}{2G_5}
    \int_0^\epsilon d\rho \tilde{F}''}.
    \label{eq4.3.18}
\end{align}
Substituting the expansion of $r(\rho)$,  $r(\epsilon) \simeq \sqrt{\alpha}+\epsilon r^{\prime}(0)$, in the last line, we obtain a simpler expression as
\begin{align}
    -\frac{1}{16\pi G_5}\int_{\mathscr{B}}\mathcal{R}\sqrt{g} d^5x
    &\simeq
    \lim_{\epsilon\rightarrow0}\bram{
    \frac{\pi^2}{2G_5}\bras{\sqrt{\alpha}+\epsilon r'(0)}^3
    \int_0^\epsilon d\rho\tilde{F}''} \notag
    \\
    &=
    \lim_{\epsilon\rightarrow0}\bras{
    \frac{\pi^2 \alpha^{\frac32}}{2G_5}
    \brab{\tilde{F}'(\epsilon)-\tilde{F}'(0)}} \notag
    \\
    &=
    \frac{\pi^2 \alpha^{\frac32}}{2G_5}
    \brab{\bras{1-\delta}-1} \notag
    \\
    &=
    -\frac{\pi^2 \alpha^{\frac32}}{2G_5}\delta.
\label{eq4.3.20}
\end{align}
This is the on-shell contribution from the EH term.

The next factor is the GHY term. Since the nondynamical term is
\begin{align}
        \oint_{\partial\mathscr{B}}
        \mathcal{K}_{0}\sqrt{\Tilde{g}}d^4\tilde{x}
        &=
        \lim_{r_s\rightarrow{}\sqrt{\alpha}}\int_{\Sigma_{r_s}}
        \mathscr{K}_0\sqrt{\gamma}d^4z \notag
        \\&=
        \lim_{r_s\rightarrow{}\sqrt{\alpha}}
        \int_{\Sigma_{r_s}}
        \frac{3}{r}\cdot r^3\bras{1-\frac{\alpha}{r^2}}^{\frac12} \sin^2{\theta_1}\sin{\theta_2}d^4z \notag
        \\&=
        0,
    \label{eq4.3.21}
\end{align}
from \eq{4.2.9}, we only need to focus on the boundary term due to $\mathscr{K}$. Using \eq{4.3.9}, we have\footnote{The coefficient in the second term differs from the calculation in \cite{Gregory:2013hja} by factor 2, but this does not affect the final results.}
\begin{align}
    \mathscr{K}
    =
    \frac{\tilde{F}'}{\tilde{F}}
    +\frac{3r'}{r}.
    \label{eq4.3.22}
\end{align}
Only the first term does affect the final result in the limit of $\epsilon\rightarrow{}0$, so the boundary integral is as follows
\begin{align}
        \frac{1}{8\pi G_5}\int_{\partial\mathscr{B}}\mathcal{K}\sqrt{\Tilde{g}}d^4\Tilde{x}
        &=
        \frac{1}{8\pi G_5}\int_{\Sigma_{\epsilon}}\mathscr{K}\sqrt{\gamma}d^4z \notag
        \\
        &=
        \lim_{\epsilon\rightarrow0}\bras{
        \frac{4\pi^3}{8\pi G_5}
        \tilde{F}(\epsilon)r(\epsilon)^3
        \mathscr{K}} \notag
        \\
        &=
        \lim_{\epsilon\rightarrow0}\bram{
        \frac{\pi^2}{2G_5}
        \tilde{F}(\epsilon)r(\epsilon)^3
        \bras{\frac{\tilde{F}'(\epsilon)}{\tilde{F}(\epsilon)}}} \notag
        \\
        &=
        \lim_{\epsilon\rightarrow0}\bras{
        \frac{\pi^2 r(\epsilon)^3}{2G_5}\tilde{F}'(\epsilon)}.
    \label{eq4.3.23}
\end{align}
Note that as the orientation of the positiveward unit normal vector and the outward normal vector coincide, the sign of the integral does not change in the first line. Using \eq{4.3.12} and the expansion of $r(\rho)$, we obtain
\begin{align}
        \frac{1}{8\pi G_5}\oint_{\partial\mathscr{B}}\mathcal{K}\sqrt{\Tilde{g}}d^4\Tilde{x}
        \simeq
        \lim_{\epsilon\rightarrow0}\bram{
        \frac{\pi^2 }{2G_5}\bras{\sqrt{\alpha}+\epsilon r'(0)}^3\tilde{F}'(\epsilon)}
	= \frac{\pi^2 \alpha^{\frac32}}{2G_5}(1-\delta).
    \label{eq4.3.24}
\end{align}
Therefore, the on-shell value of the GHY term in the limit of $\epsilon\rightarrow{}0$ is 
\begin{align}
        -\frac{1}{8 \pi G_5} \oint_{\partial \mathscr{B}} \bras{\mathcal{K}-\mathcal{K}_{0}}\sqrt{\Tilde{g}}d^4\Tilde{x}
        \simeq
        -\frac{\pi^2 \alpha^{\frac32}}{2G_5}(1-\delta).
    \label{eq4.3.25}
\end{align}

Now, we have obtained all the necessary factors. Substituting \eq{4.3.20} and \eq{4.3.25} into \eq{4.1.5}, we obtain
\begin{align}
        I_{\mathscr{B}}
        &\simeq
        -\frac{\pi^2 \alpha^{\frac32}}{2G_5}\delta
        -\frac{\pi^2 \alpha^{\frac32}}{2G_5}(1-\delta) \notag
        \\&=
        -\frac{\pi\alpha^{\frac32}}{4G_4R}.
    \label{eq4.3.26}
\end{align}
This is the very contribution to the on-shell action from the singularity.

\subsection{Total bounce action for the singular instanton}
\label{totalBounce}

Together with expressions we discussed so far, we will derive the total bounce action $ B_{\text{singular}}$. Substituting \eq{4.2.15} and \eq{4.3.26} into the Euclidean action \eq{4.1.3}, we can calculate as
\begin{align}
        B_{\text{singular}}
        \simeq
        I-I_0
        =
        \frac{\pi\alpha}{8G_4}\bras{3-\frac{2\sqrt{\alpha}}{R}}.
    \label{eq4.4.3}
\end{align}
Taking the ratio to Witten's original bounce action, we get
\begin{align}
    \frac{B_{\text{singular}}}{B}
    =
    \frac{\alpha}{R^2}\bras{3-\frac{2\sqrt{\alpha}}{R}}.
    \label{eq4.4.4}
\end{align}
\begin{figure}[tb]
\centering
\includegraphics[width=100mm]{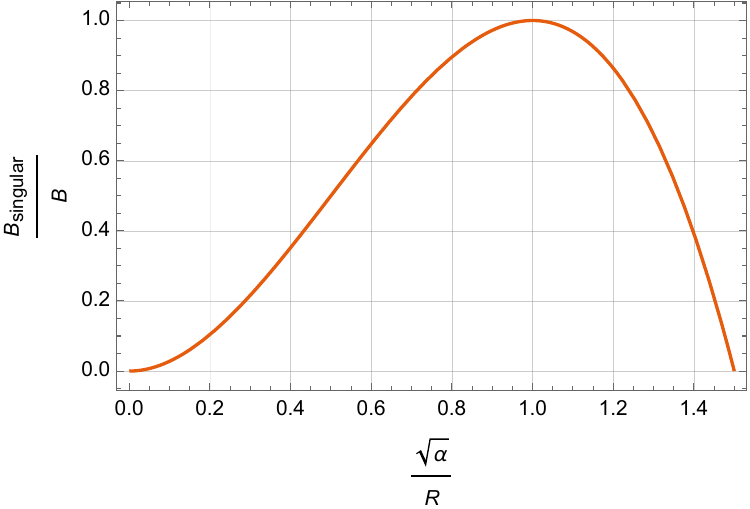}
\caption{$\sqrt{\alpha}$ dependence of the ratio between two bounce actions, the one for singular BoN instanton solution and the other for the nonsingular solution. The ratio is nondimensionalized by $\alpha/R^2$ when drawing the graph.}
\label{fig0.0}
\end{figure}
Figure \fig{0.0} illustrates a change of the ratio with respect to $\frac{\sqrt{\alpha}}{R}$. This result shows that $B_{\text{singular}}=B$ for $\sqrt{\alpha}=R$, which reproduces the result for the smooth instanton. This means that our analysis is consistent with Witten's result\cite{Witten:1981gj}. In addition, since the decay rate $\Gamma \propto e^{-B/\hbar}$ increases with smaller bounce action, \eq{4.4.4} indicates that the singularity promotes the decay of the Kaluza-Klein vacuum. The parameter $\delta$ for the deficit angle given by \eq{4.3.8.2} indicates that the larger the deficit angle, the more the vacuum decay is promoted. Here, $0<\alpha<R^2$ is the region where the deficit angle is negative, which is physically considered to be the region where the cone of singularity is turned over.

Finally, we would like to comment on the validity of our analysis for the singular instanton. In the region where the value of the bounce action is negative, the decay rate exponentially diverges, and the semiclassical approximation breaks down. Moreover, since the accuracy of the semiclassical approximation is better the larger the Euclidean action is, the accuracy is considered to become worse the closer $B_{\text{singular}}$ is to $0$. The presence of asymmetry around $\sqrt{\alpha}=R$, even in the part of the graph where the bounce action is positive, is thought to result from the poor accuracy of the approximation.

\section{Thermodynamical interpretation}
\label{BoN-thermo}

Now that we have found that the singularity promotes decay, let us consider this phenomenon from a thermodynamic point of view. It is well known that Euclidean gravitational theory can be interpreted thermodynamically\cite{Gibbons:1976ue, York:1986it, Braden:1990hw}. In the semiclassical analysis, we can also reproduce the bounce action via thermodynamic functions\cite{Dowker:1995gb,Dowker:1995sg}. Since the Euclidean action is the Helmholtz free energy divided by the Hawking temperature, it would be desirable to focus on calculating the free energy \cite{Landau:1980mil}. The  free energy can be expressed by the ADM energy $E$ and the black hole entropy $S$ as follows
\begin{align}
W=E-TS, 
\label{HermholtzFree}
\end{align}
where the temperature is the reciprocal of the period in the $\phi$ direction i.e. $T={1\over 2\pi R}$.This periodicity is constant regardless of singularity. For an asymptotically flat and geometrically complete manifold, the ADM energy is well defined by a boundary integral on the codimension-two surface\cite{Hawking:1995fd}. Similarly, entropy can be calculated by the Beckenstein-Hawking formula\cite{Bekenstein:1973ur,Bekenstein:1974ax,Hawking:1975vcx}. In fact, if the instanton solution satisfies the smoothness condition, these two quantities are calculated as 
\begin{gather}
M = \frac{3\pi R^{2}}{8G_{5}}, \label{ADMmass-witten}\\
S = \frac{A}{4G_{5}} = \frac{2\pi^{2}R^{3}}{4G_{5}}, \label{entropy-witten}
\end{gather}
and thus substituting these into \eqref{HermholtzFree} we obtain the free energy as in 
\begin{align}
W = \frac{\pi R^{2}}{8G_{5}} = \frac{\pi R^{2}}{8G_{4}\cdot 2\pi R},
\end{align}
where we used $G_5=2\pi R G_4$ in the second deformation. Therefore, by dividing it by the Hawking temperature, we can correctly reproduce the bounce action as follows
\begin{align}
B = \frac{W}{T} = \frac{\pi R^{2}}{8G_{4}}.
\end{align}

If there are conical singularities, we must carefully derive the energy and the entropy, considering the appropriate regularization. First, we will review the derivation of the ADM energy. The ADM energy for asymptotically flat and smooth spacetimes is defined by choosing the lapse function and the shift function as $N=1$, $N^a=0$ and imposing the Hamiltonian constraint,
\begin{align}
K^{a b} K_{a b}-K^2-{ }^4 R
=
0,
\label{HamiltonianConstraint}
\end{align}
 in a given Hamiltonian below \footnote{To derive the Hamiltonian, we need to evaluate the Ricci scalar of the EH term on the hypersurface, from which a new boundary term emerges. In this case, the Gauss theorem is used in the general discussion, but the theorem can not be applied to singular manifolds. However, if a smooth cap replaces the vicinity of the singularity by the conical deficit regularization, the Gauss theorem can be applied as usual.}
\begin{align}
\begin{aligned}
(16 \pi G_5) H
= & 
\int_{\Sigma_t}\left[N\left(K^{a b} K_{a b}-K^2-{ }^4 R\right)-2 N_a\left(K^{a b}-K h^{a b}\right)_{\mid b}\right] \sqrt{h} \mathrm{~d}^4 y \\
&\hspace{2.5cm} -2 \oint_{S_t}\left[N\left(k-k_0\right)-N_a\left(K^{a b}-K h^{a b}\right) r_b\right] \sqrt{\sigma} \mathrm{d}^3 \theta.
\end{aligned}
\label{eqq.1}
\end{align}
Here, for manifolds with conical singularities, it is not obvious whether the scalar curvature in the constraint condition converges or not. However, we can show that \eqref{HamiltonianConstraint} holds universally, regardless of whether we perform any regularization. The positiveward unit normal vector to $\phi$-constant surface $\Sigma_{\phi}$ is given by 
\begin{align}
    \ten{n}{}{\alpha}
    =
    \frac{\partial_\alpha\phi}{|\ten{g}{\phi\phi}{}\ten{\Phi}{}{,\phi}\ten{\Phi}{}{,\phi}|^\frac12}
    =
    \frac{\partial_\alpha \phi}{|f(r)^{-1}|^\frac12}
    =
    \sqrt{f}\partial_\alpha \phi.
    \label{eqq.5}
\end{align}
Since the instanton spacetime does not depend on $\phi$, each component of the extrinsic curvature vanishes. We can also compute the four-dimensional Ricci scalar on the $\phi$-constant hypersurface as follows
\begin{align}
        {}^4R
        =
        -\frac{6r''(\rho)}{r(\rho)}
        +\frac{6(1-r'(\rho)^2)}{r(\rho)^2},
    \label{eqq.9}
\end{align}
where $\rho$ is the radial coordinate introduced in \eqref{eq4.3.1}. Since the derivatives of $r$ with respect to $\rho$ are
\begin{align}
r^{\prime} = \bras{1-\frac{\alpha}{r^2}}^{\frac12},\quad r''
    =
    \frac{\alpha}{r^3},
\end{align}
we can verify that the Ricci scalar is zero. Therefore, the constraint condition still holds even in the presence of conical singularities, and the ADM energy is defined by
\begin{align}
E = -\frac{1}{8\pi G_{5}}\oint_{S_{\phi r}}(k-k_0)\sqrt{\sigma}d^3 \theta,
\label{ADMmass}
\end{align}
where $S_{\phi r}$ denote the $(\phi,r)$-constant three surface.

Let us compute the ADM energy based on \eqref{ADMmass}. From the definition of the extrinsic curvature, we obtain its trace as 
\begin{align}
k = \frac{3\sqrt{f(r)}}{r}.
\label{extrinsic-ADM}
\end{align}
To obtain the extrinsic curvature embedded into the flat spacetime, all we have to do is simply fix $\alpha$ to zero in \eqref{extrinsic-ADM}, then we get
\begin{align}
k_{0} = \frac{3}{r}.
\label{extrinsic0-ADM}
\end{align}
In the case of the BoN instanton solution, even this simplified calculation yields the correct result. Since the difference between \eqref{extrinsic-ADM} and \eqref{extrinsic0-ADM} does not diverge near the singularity, the boundary integral clearly converges. When we adopt the conical deficit regularization denoted in the section \ref{EuclidAction_B}, contributions from two spatial boundaries arise but will cancel each other. In the end, as in the case of smooth manifolds, we only need to consider the contribution from the asymptotic boundary, and we obtain 
\begin{align}
E = \frac{3\pi \alpha}{8G_{5}}.
\end{align}
This result coincides with \eqref{ADMmass-witten} under $\sqrt{\alpha} = R$ and indicates no correction from the conical singularity to the energy, at least at the classical level.

Next, we consider the entropy. Microscopic degrees of freedom of black holes are still under discussion, and due to their unclearness, various methods have been proposed in the calculation of entropy; see \cite{Wald:1999vt} for a comprehensive review. Here, we adopt one traditional method, which defines entropy as a semiclassical approximation of the thermodynamic potential of the microcanonical ensemble\cite{Brown:1992bq}. The Euclidean action of the microcanonical ensemble is defined as a Legendre transformation of the ordinary gravitational action as follows
\begin{align}
I_{mc,{\rm E}} \equiv I_{grav,{\rm E}} -\beta\int_{\mathcal{B}_{\tau}}d^{n-2}z\left(N\Tilde{\varepsilon}-N^{a}\Tilde{j}_{a} \right),
\end{align}
where $\varepsilon$ and $j_{a}$ are the quasi-local energy and angular momentum density defined by the Brown-York tensor\cite{Brown:1992br}, and the tilde means that the volume element $\sqrt{\sigma}$ is contained. $\beta$ is the inverse temperature and holds $\beta=2\pi R$ in our case. With the above Euclidean action, black hole entropy is given by\cite{Brown:1992bq}
\begin{align}
    S= -{\Gamma}_{mc} = -I_{mc,{\rm E}} \big|_{\rm cl}.
\end{align}
We have already derived the gravitational action in \eqref{eq4.4.3}, and the subtracted term in the Legendre transformation coincides with the ADM energy, i.e., 
\begin{align}
    \beta\int_{\partial \left(\mathscr{V}-\mathscr{B} \right) }d^{n-2}z\left(N\Tilde{\varepsilon}-N^{a}\Tilde{j}_{a} \right) = \frac{3\pi \alpha \beta}{8G_{5}}, 
\end{align}
where we choose $N\rightarrow 1$, $N^{a}\rightarrow 0$. Thus, the entropy calculated by the microcanonical ensemble yields
\begin{align}
S = \frac{2\pi^{2}\alpha^{3/2}}{4G_{5}},
\end{align}
which is in perfect agreement with the Bekenstein-Hawking formula.

From the above discussions, the bounce action is given by
\begin{align}
B_{\rm singular}&={W\over T} \notag \\
&={E-TS\over T} \notag \\
&= {3\pi \alpha \over 8 G_4 } - {\pi \alpha^{3/2}\over 4 RG_4}={\pi \alpha \over 8 G_4 }\Big(3-{2\sqrt{\alpha }\over R} \Big), 
\end{align}
which agrees with the result \eq{4.4.3} obtained in the previous section. This consistency could be regarded as evidence of the correctness of the prescription used in \cite{Gregory:2013hja}.

How can we interpret the fact that conical singularities enhance instability? Let us consider shifting $\alpha$ slightly from $R^{2}$ and express the change in energy and entropy, respectively, up to the second order as follows
\begin{align}
\Delta E= {3  \over 16G_4 R}\Delta \alpha\ , \qquad \Delta S= {3\pi \over 8G_4}\Delta \alpha+{3\pi \over 32 G_4 R^2}\Delta \alpha^2.
\end{align}
Substituting these into thermodynamic definition yields
\begin{align}
B_{\rm singular}&=B+{\Delta E \over T}-\Delta S \nonumber\\
&\simeq B+ {3\pi   \over 8G_4 }\Delta \alpha - {3\pi \over 8G_4}\Delta \alpha-{3\pi \over 32 G_4 R^2}\Delta \alpha^2 \nonumber \\
&= B-{3\pi \over 32 G_4 R^2}\Delta \alpha^2.
\end{align}
The change in the energy is canceled with the first-order change in the entropy, and then only the second-order change in the entropy remains. This second-order change is always a negative quantity, indicating that the change in the horizon position is a factor that lowers the bounce action. In other words, the energy would be larger, but the entropy would be larger than that, lowering the free energy. This is the thermodynamic interpretation that singularities in instanton promote vacuum decay.

\section{Discussions and Summary}
\label{DiscussionSummary}

In this paper, we investigated the vacuum decay of the five-dimensional Kaluza-Klein spacetime using a singular bubble of nothing solution. The bounce action, which governs the decay rate, becomes smaller than that of Witten's solution without singularity, thereby increasing the decay rate. This consequence suggests that instantons with singularities play a dominant role in higher-dimensional theories with compact internal spaces and are expected to play an essential role in the ongoing discussions of the Swampland conjectures. The Euclideanized theory allows for a thermodynamic interpretation, which we also explored. The contributions from singularities increase the internal energy contribution, but this is outweighed by an increase in entropy, leading to a net decrease in free energy, thereby enhancing vacuum decay.

Given the thermodynamical analysis, what number of states does entropy here come from? One possibility is the way to regularize the singularity. The authors of \cite{Gregory:2013hja} showed that contributions of the singularity in calculating on-shell action do not depend on the scheme once the spacetime is smoothly regularized. This means that if we do not take $\epsilon \rightarrow 0$ limit and remain $\epsilon$ finite, there will be many states with slightly different profiles in the vicinity of the singularity. This number of states could be interpreted as contributing to the increase in entropy.

Discussions on variants of the bubble of nothing solution within the framework of string theory have already been made \cite{Horowitz:2007pr,Ooguri:2017njy}. We believe that instantons with singularities also have a dominant contribution in the context of string theory. We will explore this direction further in a separate publication. Moreover, this paper deals with the transition between two vacua by focusing on gravity without explicitly introducing scalar fields. To handle more realistic models, the inclusion of such scalar fields is necessary. A study shown in \cite{Blanco-Pillado:2023aom} can be helpful in this direction, and attempting to reproduce what we have done with more concrete models would be helpful for a better understanding of the contributions from singularities.

Finally, let us comment on negative modes around the classical solution. In this paper, we simply assumed that the analysis of the negative modes discussed by Witten is also applicable here, since we basically used the same solution. As Coleman has claimed, only one negative mode is essential for the decay to be worked. The treatment of singularities in this paper might introduce new negative modes, in fact previous studies \cite{Lee:2014uza,Gregory:2018bdt} have shown that there are more than one negative mode around gravitational instantons. Hence it is required that more careful treatment for the negative mode has to be done. We will revisit on this issue in a future work.

\section*{Acknowledgments}

This work is supported by Grant-in-Aid for Scientific Research from the Ministry of Education, Culture, Sports, Science and Technology, Japan (JP20K03932). 

\appendix

\appendix
\section{Notation}
\label{notation}

We summarize the notation for this paper here. Greek indices ($\alpha,\beta,\gamma,\cdots$) denote five-metric, Latin indices ($i,j,k,\cdots$) denote four-metric, and capital letters ($A,B,C,\cdots$) denote three-metric. $\Omega_n$ is a $n$-dimensional angular coordinates i.e. $\Omega_n=(\theta_1,\theta_2,\theta_3,\cdots,\theta_{n-1},\theta_n$). When representing an arbitrary coordinate constant surface, we assign coordinates that take constant values to $\Sigma$ as subscripts. For example, a $\phi$-constant surface is represented as $\Sigma_{\phi}$. When dealing with other geometric quantities on surfaces, we shall follow the notation in Table \tab{1.1}. Also, unit vectors $n^{\alpha}$ and $r^{\alpha}$ are assumed to point in the positive direction.

\begin{table}[h]
  \caption{Geometric quantities of surfaces appeared in this paper.}
  \label{tab1.1}
  \centering
\begin{tabular}{ccccc} \hline\hline
   Surface & arbitrary surface & $\Sigma_\phi$ & $\Sigma_r$($\Sigma_\rho$) & $S_{\phi r}$ \\ \hline
   Unit normal vector & $\Tilde{n}^\alpha$\ (outward) & $n^\alpha$ & $r^\alpha$ & $r^\alpha$ \\
   Subscript & $a,b,\cdots$ & $a,b,\cdots=(r,\Omega_3)$ & $i,j,\cdots=(\phi,\Omega_3)$ & $A,B,\cdots=(\Omega_3)$ \\
   Coordinates & $\tilde{x}^{a}$ & $y^a$ & $z^i$ & $\theta^A$ \\
   Induced metric &$\tilde{g}_{ab}$& $h_{ab}$ & $\gamma_{ij}$ & $\sigma_{AB}$ \\
   Extrinsic curvature & $\mathcal{K}_{ab}$ & $K_{ab}$ & $\mathscr{K}_{ij}$ & $k_{AB}$ \\ \hline\hline
 \end{tabular}
 \end{table}

\section{Derivation of Euclidean action via ADM decomposition}
\label{EuclidAction-3+1}

In section \ref{EuclidAction_V/B}, we derived the Euclidean action of $\mathscr{V}-\mathscr{B}$ directly from the definition of the gravitational action, while the authors of \cite{Gregory:2013hja} used the ADM decomposition\cite{Hawking:1995fd} to evaluate this value. Here, we also compute $I_{\mathscr{V}-\mathscr{B}}$ using the ADM decomposition and show that our computation is consistent with their argument.

As shown in the beginning of the section \ref{Quick review of bubble of nothing solution}, the gravitational action is given by
\begin{align}
    I_{\mathscr{V}-\mathscr{B}}
    =
    -\frac{1}{16 \pi G_5} \int_{\mathscr{V}-\mathscr{B}} \mathcal{R}\sqrt{g} d^5x
    -\frac{1}{8 \pi G_5} \oint_{\partial \bras{\mathscr{V}-\mathscr{B}}} \bras{\mathcal{K}-\mathcal{K}_{0}}\sqrt{\Tilde{g}}d^4\Tilde{x}.
    \label{eqJ.1}
\end{align}
The ADM decomposition by the $\phi$-constant hypersurface $\Sigma_{\phi}$ yields the following expression\cite{Hawking:1995fd}
\begin{align}
I_{\mathscr{V}-\mathscr{B}}&=-\frac{1}{16\pi G_{5}}\int^{2\pi R}_{0}d\phi \left[\int_{\Sigma_{\phi}}\left({}^{(4)}\partial_{\phi}g_{ij}\pi^{ij}-N\mathcal{H}-N^{i}\mathcal{H}_{i}\right)N\sqrt{h}d^{4}y -2\oint_{S_{\phi r}}(k-\mathscr{K}_{0})N\sqrt{\sigma}d^{3}\theta \right].
\end{align}
We adopt here the notation in the original  $r$-constant surface for the nondynamical term, but we can confirm that this agrees with $k_{0}$ by an explicit calculation. As the volume integral disappears due to the constraint condition, $\mathcal{H}=\mathcal{H}_{i}=0$, and the symmetry of spacetime with respect to $\phi$, we only need to focus on the boundary integral.\footnote{See also discussions in section \ref{BoN-thermo}.} As mentioned in section \ref{EuclidAction_V/B}, $\mathscr{V}-\mathscr{B}$ includes two boundary surfaces, the spatial boundary and the asymptotic boundary, so we must evaluate the integral for each boundary. That is to say,
\begin{align}
        I_{\mathscr{V}-\mathscr{B}}
        &=
        \frac{1}{8 \pi G_5} \int_0^{2\pi R}d\phi 
        \bram{\lim_{r_s\rightarrow\sqrt{\alpha}}\oint_{S_{\phi r_s}}(k-\mathscr{K}_0)N\sqrt{\sigma}d^3\theta
        -\lim_{r_\infty\rightarrow\infty}\oint_{S_{\phi r_\infty}}(k-\mathscr{K}_0)N\sqrt{\sigma}d^3\theta}. \label{action-V/B-ADMdecom}
\end{align}
The trace of the extrinsic curvature for $(\phi,r)$-constant surface is computed as
\begin{align}
k = \frac{3\sqrt{f(r)}}{r} ,\quad f(r)=1-\frac{\alpha}{r^{2}}.
\end{align}
We have already derived the trace of the extrinsic curvature of the surface embedded into flat spacetime in \eq{3.2.16}: 
\begin{align}
\mathscr{K}_{0} = \frac{3}{r}.
\end{align}
The volume element is given by 
\begin{align}
    N\sqrt{\sigma}
    =
    \sqrt{f(r)}\ r^3\sin^2{\theta_1}\sin{\theta_2}.
    \label{eqJ.19}
\end{align}
Substituting all of the above into \eqref{action-V/B-ADMdecom}, we obtain the Euclidean action as follows
\begin{align}
I_{\mathscr{V}-\mathscr{B}}
	& = \frac{2\pi R}{8\pi G_{5}} \left\{
        \lim_{r_s\rightarrow\sqrt{\alpha}}\oint_{S_{\phi r_s}}\bras{\frac{3\sqrt{f(r)}}{r}-\frac{3}{r}}\sqrt{f(r)}\ r^3\sin^2{\theta_1}\sin{\theta_2}d^3\theta \right. \notag \\
        & \left. \hspace{2.5cm}-\lim_{r_\infty\rightarrow\infty}\oint_{S_{\phi r_\infty}}\bras{\frac{3\sqrt{f(r)}}{r}-\frac{3}{r}}\sqrt{f(r)}\ r^3\sin^2{\theta_1}\sin{\theta_2}d^3\theta
        \right\} \notag \\
        & = \frac{3\cdot 2\pi^{2}}{8\pi G_{4}}\left\{\lim_{r_s\rightarrow\sqrt{\alpha}} \left(f(r) - \sqrt{f(r)} \right)r^{2} - \lim_{r_\infty\rightarrow\infty}\left(f(r) - \sqrt{f(r)} \right)r^{2} \right\} \notag \\
        & = \frac{3\cdot 2\pi^{2}}{8\pi G_{4}}\left\{0 - \left(-\frac{\alpha}{2} \right) \right\} \notag \\
        & = \frac{3\pi \alpha}{8G_{4}}. \label{onshellAction-V/B-ADM}
\end{align}
This result coincides with \eq{4.2.15} in section \ref{EuclidAction_V/B}, which we calculated without the ADM decomposition.

In the above calculation, the boundary integral has a finite value, which appears to be a different result from \cite{Gregory:2013hja}. However, this discrepancy is due to the difference that they considered a compact spacetime (de Sitter-Schwarzshild spacetime), whereas we considered a noncompact spacetime and have an asymptotic boundary at infinity. Therefore, our result is consistent with their argument.

%


\begin{thebibliography}{99}

\bibitem{Coleman:1977py}
S.~R.~Coleman,
Phys. Rev. D \textbf{15}, 2929-2936 (1977)
[erratum: Phys. Rev. D \textbf{16}, 1248 (1977)].

\bibitem{Callan:1977pt}
C.~G.~Callan, Jr. and S.~R.~Coleman,
Phys. Rev. D \textbf{16}, 1762-1768 (1977).

\bibitem{Coleman:1980aw}
S.~R.~Coleman and F.~De Luccia,
Phys. Rev. D \textbf{21}, 3305 (1980).

\bibitem{Weinberg}
Erick J. Weinberg,
``Classical Solutions in Quantum Field Theory,'' CAMBRIDGE, 2012.

\bibitem{Hawking:1998bn}
S.~W.~Hawking and N.~Turok,
Phys. Lett. B \textbf{425}, 25-32 (1998)
[arXiv:hep-th/9802030].

\bibitem{Turok:1998he}
N.~Turok and S.~W.~Hawking,
Phys. Lett. B \textbf{432}, 271-278 (1998)
[arXiv:hep-th/9803156].

\bibitem{Unruh:1998wc}
W.~G.~Unruh,
[arXiv:gr-qc/9803050].

\bibitem{Vilenkin:1998pp}
A.~Vilenkin,
Phys. Rev. D \textbf{57}, 7069-7070 (1998)
[arXiv:hep-th/9803084].

\bibitem{Garriga:1998tm}
J.~Garriga,
Phys. Rev. D \textbf{61}, 047301 (2000)
[arXiv:hep-th/9803210].

\bibitem{Garriga:1998ri}
J.~Garriga,
[arXiv:hep-th/9804106].

\bibitem{Garriga:1998he}
J.~Garriga, X.~Montes, M.~Sasaki and T.~Tanaka,
Nucl. Phys. B \textbf{551}, 317-373 (1999)
[arXiv:astro-ph/9811257].

\bibitem{Hashimoto:2000zk}
M.~Hashimoto,
Phys. Rev. D \textbf{63}, 043510 (2001)
[arXiv:hep-th/0005102].

\bibitem{Gregory:2013hja}
R.~Gregory, I.~G.~Moss and B.~Withers,
JHEP \textbf{03}, 081 (2014)
[arXiv:hep-th/1401.0017].

\bibitem{Fursaev:1995ef}
D.~V.~Fursaev and S.~N.~Solodukhin,
Phys. Rev. D \textbf{52}, 2133-2143 (1995)
[arXiv:hep-th/9501127].

\bibitem{Schon:1979rg}
R.~Schon and S.~T.~Yau,
Commun. Math. Phys. \textbf{65}, 45-76 (1979).

\bibitem{Schon:1981vd}
R.~Schon and S.~T.~Yau,
Commun. Math. Phys. \textbf{79}, 231-260 (1981).

\bibitem{Witten:1981mf}
E.~Witten,
Commun. Math. Phys. \textbf{80}, 381 (1981).

\bibitem{Witten:1981gj}
E.~Witten,
Nucl. Phys. B \textbf{195}, 481-492 (1982).

\bibitem{Horowitz:2007pr}
G.~T.~Horowitz, J.~Orgera and J.~Polchinski,
Phys. Rev. D \textbf{77}, 024004 (2008)
[arXiv:hep-th/0709.4262].

\bibitem{Blanco-Pillado:2016xvf}
J.~J.~Blanco-Pillado, B.~Shlaer, K.~Sousa and J.~Urrestilla,
JCAP \textbf{10}, 002 (2016)
[arXiv:1606.03095 [hep-th]].

\bibitem{Acharya:2019mcu}
B.~S.~Acharya,
JHEP \textbf{08}, 128 (2020)
[arXiv:1906.06886 [hep-th]].

\bibitem{GarciaEtxebarria:2020xsr}
I.~Garc\'\i{}a Etxebarria, M.~Montero, K.~Sousa and I.~Valenzuela,
JHEP \textbf{12}, 032 (2020)
[arXiv:2005.06494 [hep-th]].

\bibitem{Vafa:2005ui}
C.~Vafa,
[arXiv:hep-th/0509212 [hep-th]].

\bibitem{Ooguri:2006in}
H.~Ooguri and C.~Vafa,
Nucl. Phys. B \textbf{766}, 21-33 (2007)
[arXiv:hep-th/0605264 [hep-th]].

\bibitem{Ooguri:2016pdq}
H.~Ooguri and C.~Vafa,
Adv. Theor. Math. Phys. \textbf{21}, 1787-1801 (2017)
[arXiv:1610.01533 [hep-th]].

\bibitem{Obied:2018sgi}
G.~Obied, H.~Ooguri, L.~Spodyneiko and C.~Vafa,
[arXiv:1806.08362 [hep-th]].
 
\bibitem{Garg:2018reu}
S.~K.~Garg and C.~Krishnan,
JHEP \textbf{11}, 075 (2019)
[arXiv:1807.05193 [hep-th]].
 
\bibitem{Ooguri:2018wrx}
H.~Ooguri, E.~Palti, G.~Shiu and C.~Vafa,
Phys. Lett. B \textbf{788}, 180-184 (2019)
[arXiv:1810.05506 [hep-th]].

\bibitem{Lust:2019zwm}
D.~Lust, E.~Palti and C.~Vafa,
Phys. Lett. B \textbf{797}, 134867 (2019)
[arXiv:1906.05225 [hep-th]].

\bibitem{Palti:2019pca}
E.~Palti,
Fortsch. Phys. \textbf{67}, no.6, 1900037 (2019)
[arXiv:1903.06239 [hep-th]].

\bibitem{Brennan:2017rbf}
T.~D.~Brennan, F.~Carta and C.~Vafa,
PoS \textbf{TASI2017}, 015 (2017)
[arXiv:1711.00864 [hep-th]].

\bibitem{Danielsson:2018ztv}
U.~H.~Danielsson and T.~Van Riet,
Int. J. Mod. Phys. D \textbf{27}, no.12, 1830007 (2018)
[arXiv:1804.01120 [hep-th]].

\bibitem{Akrami:2018ylq}
Y.~Akrami, R.~Kallosh, A.~Linde and V.~Vardanyan,
Fortsch. Phys. \textbf{67}, no.1-2, 1800075 (2019)
[arXiv:1808.09440 [hep-th]].

\bibitem{Grana:2021zvf}
M.~Gra\~na and A.~Herr\'aez,
Universe \textbf{7}, no.8, 273 (2021)
[arXiv:2107.00087 [hep-th]].

\bibitem{vanBeest:2021lhn}
M.~van Beest, J.~Calder\'on-Infante, D.~Mirfendereski and I.~Valenzuela,
Phys. Rept. \textbf{989}, 1-50 (2022)
[arXiv:2102.01111 [hep-th]].

\bibitem{Banerjee:2018qey}
S.~Banerjee, U.~Danielsson, G.~Dibitetto, S.~Giri and M.~Schillo,
Phys. Rev. Lett. \textbf{121}, no.26, 261301 (2018)
[arXiv:1807.01570 [hep-th]].

\bibitem{Banerjee:2019fzz}
S.~Banerjee, U.~Danielsson, G.~Dibitetto, S.~Giri and M.~Schillo,
JHEP \textbf{10}, 164 (2019)
[arXiv:1907.04268 [hep-th]].

\bibitem{Banerjee:2020wix}
S.~Banerjee, U.~Danielsson and S.~Giri,
JHEP \textbf{04}, 085 (2020)
[arXiv:2001.07433 [hep-th]].

\bibitem{York:1972sj}
J.~W.~York, Jr.,
Phys. Rev. Lett. \textbf{28}, 1082-1085 (1972).

\bibitem{Gibbons:1976ue}
G.~W.~Gibbons and S.~W.~Hawking,
Phys. Rev. D \textbf{15}, 2752-2756 (1977).

\bibitem{Brown:2014rka}
A.~R.~Brown,
Phys. Rev. D \textbf{90}, no.10, 104017 (2014)
[arXiv:1408.5903 [hep-th]].

\bibitem{Dowker:1995gb}
F.~Dowker, J.~P.~Gauntlett, G.~W.~Gibbons and G.~T.~Horowitz,
Phys. Rev. D \textbf{52}, 6929-6940 (1995)
[arXiv:hep-th/9507143 [hep-th]].

\bibitem{Adams:2001sv}
A.~Adams, J.~Polchinski and E.~Silverstein,
JHEP \textbf{10}, 029 (2001)
[arXiv:hep-th/0108075 [hep-th]].

\bibitem{York:1986it}
J.~W.~York, Jr.,
Phys. Rev. D \textbf{33}, 2092-2099 (1986).

\bibitem{Braden:1990hw}
H.~W.~Braden, J.~D.~Brown, B.~F.~Whiting and J.~W.~York, Jr.,
Phys. Rev. D \textbf{42}, 3376-3385 (1990).

\bibitem{Dowker:1995sg}
F.~Dowker, J.~P.~Gauntlett, G.~W.~Gibbons and G.~T.~Horowitz,
Phys. Rev. D \textbf{53}, 7115-7128 (1996)
[arXiv:hep-th/9512154 [hep-th]].

\bibitem{Landau:1980mil}
L.~D.~Landau and E.~M.~Lifshitz,
``Statistical Physics, Part 1,''
Butterworth-Heinemann, 1980,
ISBN 978-0-7506-3372-7.

\bibitem{Hawking:1995fd}
S.~W.~Hawking and G.~T.~Horowitz,
Class. Quant. Grav. \textbf{13}, 1487-1498 (1996)
[arXiv:gr-qc/9501014 [gr-qc]].

\bibitem{Bekenstein:1973ur}
J.~D.~Bekenstein,
Phys. Rev. D \textbf{7}, 2333-2346 (1973).

\bibitem{Bekenstein:1974ax}
J.~D.~Bekenstein,
Phys. Rev. D \textbf{9}, 3292-3300 (1974).

\bibitem{Hawking:1975vcx}
S.~W.~Hawking,
Commun. Math. Phys. \textbf{43}, 199-220 (1975)
[erratum: Commun. Math. Phys. \textbf{46}, 206 (1976)].

\bibitem{Wald:1999vt}
R.~M.~Wald,
Living Rev. Rel. \textbf{4}, 6 (2001)
[arXiv:gr-qc/9912119 [gr-qc]].

\bibitem{Brown:1992bq}
J.~D.~Brown and J.~W.~York, Jr.,
Phys. Rev. D \textbf{47}, 1420-1431 (1993)
[arXiv:gr-qc/9209014 [gr-qc]].

\bibitem{Brown:1992br}
J.~D.~Brown and J.~W.~York, Jr.,
Phys. Rev. D \textbf{47}, 1407-1419 (1993)
[arXiv:gr-qc/9209012 [gr-qc]].

\bibitem{Ooguri:2017njy}
H.~Ooguri and L.~Spodyneiko,
Phys. Rev. D \textbf{96}, no.2, 026016 (2017)
[arXiv:1703.03105 [hep-th]].

\bibitem{Blanco-Pillado:2023aom}
J.~J.~Blanco-Pillado, J.~R.~Espinosa, J.~Huertas and K.~Sousa,
[arXiv:2312.00133 [hep-th]].

\bibitem{Lee:2014uza}
H.~Lee and E.~J.~Weinberg,
Phys. Rev. D \textbf{90}, no.12, 124002 (2014)
[arXiv:1408.6547 [hep-th]].

\bibitem{Gregory:2018bdt}
R.~Gregory, K.~M.~Marshall, F.~Michel and I.~G.~Moss,
Phys. Rev. D \textbf{98}, no.8, 085017 (2018)
[arXiv:1808.02305 [hep-th]].

\end{thebibliography}
\end{document}